\newcommand{\ie}{\mbox{\itshape i.e.}}
\newcommand{\eg}{\mbox{\itshape e.g.}}
\newcommand{\mrad}{\ensuremath{\mathrm{mrad}}}
\newcommand{\murad}{\ensuremath{\mathrm{\mu rad}}}
\newcommand{\mum}{\ensuremath{\mathrm{\mu m}}}
\newcommand{\cm}{\ensuremath{\mathrm{cm}}}

\newcommand{\R}{\ensuremath{R}}
\newcommand{\rvar}{\ensuremath{r}}
\newcommand{\x}{\ensuremath{\mathrm{x}}}
\newcommand{\y}{\ensuremath{\mathrm{y}}}
\newcommand{\z}{\ensuremath{\mathrm{z}}}
\newcommand{\X}{\ensuremath{\mathrm{X}}}
\newcommand{\Y}{\ensuremath{\mathrm{Y}}}
\newcommand{\Z}{\ensuremath{\mathrm{Z}}}

\documentclass{elsart}
\usepackage{natbib}
\usepackage{amssymb}
\usepackage{graphicx}
\usepackage{glas}
\begin{document}

\begin{titlepage}{GLAS-PPE/2008-06}{10$^{\underline{\rm{th}}}$ July 2008}

\title{Alignment procedure \\of the LHCb Vertex Detector}

\author{S.~Viret, C.~Parkes, M.~Gersabeck\\
\\
Department of Physics and Astronomy, University of Glasgow\\
University Avenue, Glasgow, G12 8QQ, United Kingdom}

\begin{abstract}
LHCb is one of the four main experiments of the Large Hadron Collider
(LHC) project, which will start at CERN in 2008. The experiment is
primarily dedicated to B-Physics and hence requires precise vertex
reconstruction. The silicon vertex locator (VELO) has a single hit
precision of better than $10~\mum$ and is used both off-line and in
the trigger. These requirements place strict constraints on its
alignment. Additional challenges for the alignment arise from the
detector being retracted between each fill of the LHC and from its
unique circular disc r/$\phi$ strip geometry. This paper describes
the track based software alignment procedure developed for the VELO.
The procedure is primarily based on a non-iterative method using a
matrix inversion technique. The procedure is demonstrated with
simulated events to be fast, robust and to achieve a suitable alignment
precision.

\end{abstract}

\newpage
\end{titlepage}

\runauthor{Sebastien Viret}
\begin{frontmatter}
\title{Alignment procedure \\of the LHCb Vertex Detector}
\author{S.~Viret, C.~Parkes, M.~Gersabeck\\
\\
Department of Physics and Astronomy, University of Glasgow\\
University Avenue, Glasgow, G12 8QQ, United Kingdom}

\begin{abstract}
LHCb is one of the four main experiments of the Large Hadron Collider
(LHC) project, which will start at CERN in 2008. The experiment is
primarily dedicated to B-Physics and hence requires precise vertex
reconstruction. The silicon vertex locator (VELO) has a single hit
precision of better than $10~\mum$ and is used both off-line and in
the trigger. These requirements place strict constraints on its
alignment. Additional challenges for the alignment arise from the
detector being retracted between each fill of the LHC and from its
unique circular disc r/$\phi$ strip geometry. This paper describes
the track based software alignment procedure developed for the VELO.
The procedure is primarily based on a non-iterative method using a
matrix inversion technique. The procedure is demonstrated with
simulated events to be fast, robust and to achieve a suitable alignment
precision.

\end{abstract}

\end{frontmatter}

\section{Introduction}

LHCb is the dedicated heavy flavour physics experiment at the LHC. The
physics goals are critically dependent on the performance of the
precision vertex locator (VELO). Whilst the intrinsic VELO sensor
single hit resolution is $5-10~\mum$ for all tracks in the
acceptance, the sensors also have to be retracted by $3~\cm$ from the
unstable LHC beams while the machine is filled. As a result of these
circumstances and the unique geometry of the detector, described in
Section~\ref{sec:context}, the VELO has particularly demanding
alignment requirements.

The VELO detector has been assembled with a high precision and a
detailed metrology has been performed. However, the only possible
method for correcting misalignments during data taking is through a
track-based software alignment procedure. A new alignment will be
performed after each re-insertion of the VELO, in order to check the
alignment and ensure optimal data taking.

The alignment procedure is based on techniques which are widely used in particle physics, a comprehensive review of particle physics alignment methods is available in Ref.~\cite{bib:Yel-07}. Two types of techniques are applied here, one is based on matrix inversion and the other on the fitting of functional forms to residual distributions. The LHCb VELO alignment is however unique due to the module and detector geometry and the frequent mechanical retraction of the detector.

The VELO alignment algorithm is divided into three phases: the relative alignment of the sensors in
each module; the relative alignment of the modules in each half of
the detector; and the relative alignment of the detector-halves. The
algorithms are described in Section~\ref{sec:align}. The results
obtained using simulated events are described in Section~\ref
{sec:MC_results}. A summary and conclusions are given in Section~\ref
{sec:conclusion}.

The performance of the first two phases of the alignment procedure
were also tested on data during a VELO beam test. The results of this
study are presented in Ref.~\cite{bib:NIM-07} and verify the
simulation results presented here.

\section{The VELO Alignment Context}
\label{sec:context}

\begin{figure}[h!]
\begin{center}
\includegraphics[width=0.55\columnwidth]{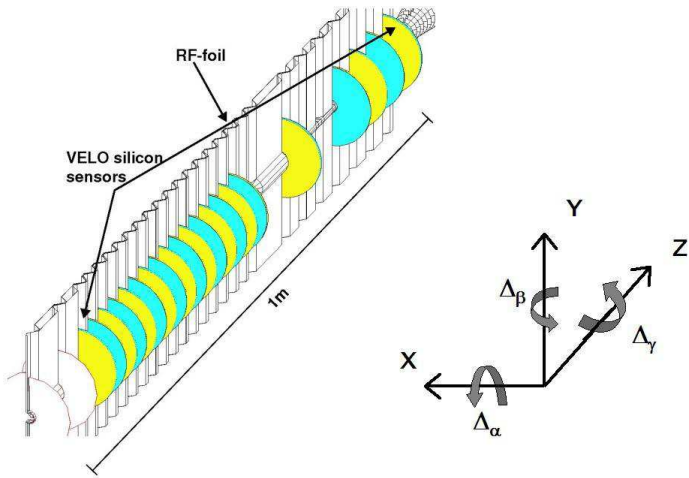}
\caption{Layout of the modules in the LHCb vertex detector (left).
The right figure describes, in the VELO reference frame, the
rotational degrees of freedom used throughout this paper. Rotations
around the different axis are applied in the following order: \Z\ ($
\Delta_{\gamma}$), \Y' ($\Delta_{\beta}$), and \X'' ($\Delta_{\alpha}
$).}
\label{fig:layout}
\end{center}
\end{figure}

The VELO consists of two retractable halves, each with 21 modules, as
shown in Fig.~\ref{fig:layout}.
Each module contains two semi-circular silicon strip sensors that
measure r (radial) and $\phi$ (azimuthal angle) co-ordinates,
respectively. The strips on the $\phi$ measuring sensor have a stereo
angle off-set. A hole up to a radius of 7~mm in the centre of the
sensors allows the two proton beams to pass through the VELO, with
their interaction point being close to the start of the detector.

The VELO sensors are operated in vacuum and separated from the
primary beam vacuum of the LHC by a thin ($250~\mum$) aluminium foil.
The two halves of the VELO will be retracted by $3~\cm$ during each
LHC fill until stable beam conditions have been established.
Further details of the VELO detector design can be found in Ref.~\cite
{bib:ReOptTDR-03}.

No magnetic field is applied in the VELO active area, and the
residual effects of the LHCb dipole magnet are small. Thus the tracks
reconstructed in the detector can be represented by straight lines in
the VELO at first order. The effect on the alignment of using straight lines, and hence neglecting the magnetic field and multiple scattering effects, in the VELO have been studied in simulation. Magnetic field effects could affect the alignment result only when a low momentum track sample with a large charge asymmetry is used. Those effects may be neglected when they are minimized by using an equal number of positively and negatively charged particles. Furthermore, one can use the LHCb Trigger Tracker~\cite{bib:TTTDR-02} information (also in an un-aligned environment) to select a track sample with equal numbers for both charges.

The successful operation of the LHCb trigger puts tight constraints
on the VELO assembly precision. The LHCb trigger system relies on
vertexing but, for speed reasons, the VELO pattern recognition
algorithm has to be performed initially in the \rvar-\z\ projection.
This requires that the strips on the R sensors accurately describe
circles around the beam position.

Moreover, remaining misalignments that are not corrected for in
software could have significant effects on the trigger performance.
For example, a rotation of 0.5~\mrad\ around the \Y-axis of one
detector can reduce the trigger efficiency by 30~\% for the $B_{s}
\rightarrow K^{+}K^{-}$ channel~\cite{bib:Dav-05}.

\section{The VELO Alignment Procedure}
\label{sec:align}

The alignment of the LHCb VELO proceeds in a number of stages:
precision assembly; metrology; and the software alignment described
in this paper.

A survey of the VELO system has been performed. The relative position
of the $\R$ and $\Phi$ sensors on a single module have been measured
to an accuracy better than $5\ \mum$, and no significant curvature of
the sensors was observed. The relative position of the VELO modules
on the VELO half bases has also been determined with a similar
precision. The survey is of particular importance for determining
degrees of freedom that can be constrained less well from tracks,
such as the relative \z-positions of the VELO modules. Moreover, the survey values have been taken as the initial VELO alignment conditions: the software alignment will be seeded with these values.

The VELO software alignment procedure is divided into three phases. These phases are motivated by and reflect the mechanical construction of the system and the different time periods on which re-alignment is anticipated.

\begin{itemize}

\item The first phase performs a relative alignment of $\R$ and $\Phi$ sensors within one VELO module~\cite{bib:Mar-07}. The $\R$ and $\Phi$ sensors are glued together onto the same hybrid, thus making their relative alignment highly stable. It is anticipated that the alignment of this structure will not need to be repeated as frequently as for the other phases. The alignment technique is based on an iterative fit of the distribution of track residuals across the surface of the silicon sensor.

\item The second phase is an internal alignment of the modules within each VELO detector-half~\cite{bib:Vir-05}. The two VELO detector halves are mechanically separate and can be retracted, partially inserted or fully inserted. The VELO halves will be retracted each fill and hence, at least as a cross-check for relative module movement, the alignment will be performed each fill.  The alignment technique is based on track residuals applying a non-iterative method using a matrix inversion technique.

\item The third phase is the relative alignment of the two halves with respect to each other~\cite{bib:Vir-07}. This will be performed each fill to cross-check and improve upon information provided from the mechanical insertion system. The alignment technique uses the same technique as the module alignment but requiring tracks that pass through both VELO halves and on a similar technique that is applied with vertex constraints.

\end{itemize}

These three alignment algorithm phases are described in more detail in the following sections.

\subsection{Relative Sensor Alignment}

A VELO module contains an $\R$ and a $\Phi$ sensor glued onto the
same hybrid, so that the sensors are back-to-back. The first step of
the alignment procedure is to determine the relative alignment of
the  $\R$ and $\Phi$ sensors. This relative alignment is already
known to high accuracy from a mechanical survey but can be further
improved by applying the following procedure to determine the
critical relative $\x$ and $\y$ translations of the sensors.

As the $\R$ and $\Phi$ sensors on each module have to be treated
separately, the residuals with respect to strip hits have to be
considered. The unique VELO $\R/\Phi$ sensor geometry means that the
linearized non-iterative approach applied to space-points in the
subsequent alignment phases is not applicable. Instead, an iterative
alignment procedure that fits the distribution of residuals as a
function of the sensor azimuthal angle has been developed. This procedure is partly inspired by the method developed for the SLD vertex detector~\cite{bib:SLD-07}.

\begin{figure}[h!]
\centering
\includegraphics[width=0.45\textwidth]{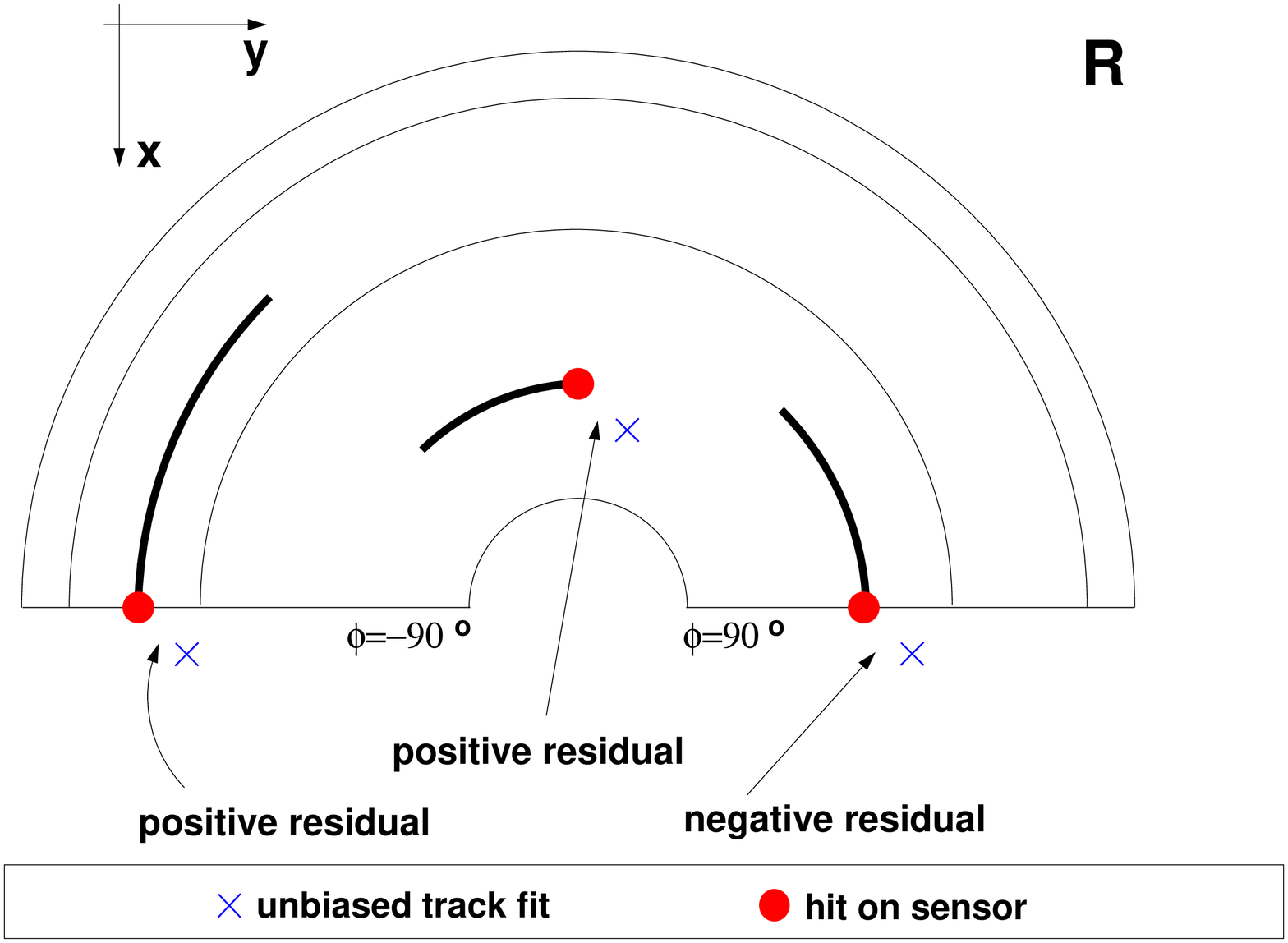}
\includegraphics[width=0.45\textwidth]{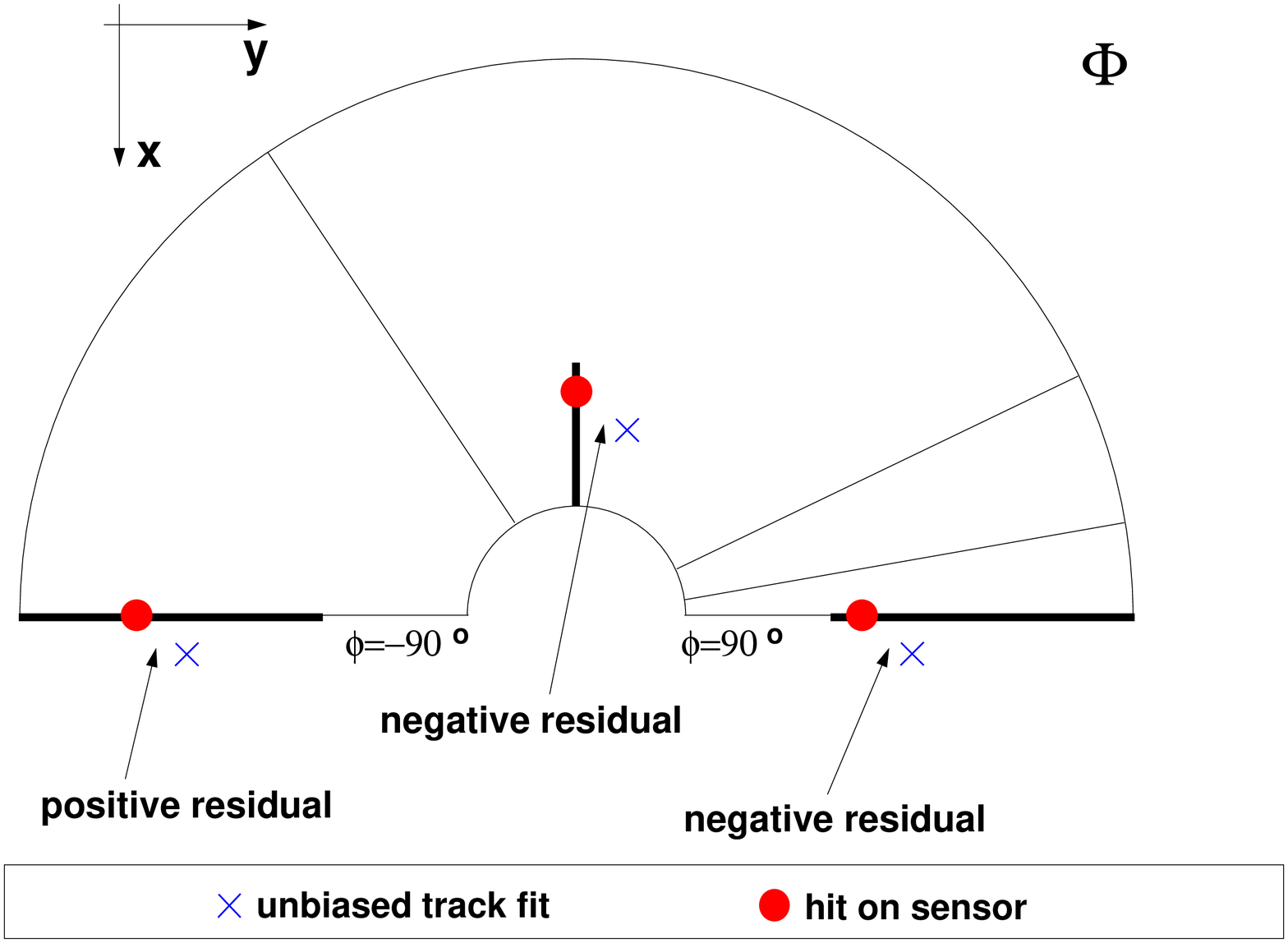}
\caption{Influence of misalignments on residuals of $R$ and $\Phi$
sensors. The misalignment shown is a translation both along the
negative $\x$ and $\y$ direction.}
\label{fig:STEP0_sketch}
\end{figure}

The observed signals on sensor strips are used to determine the best
estimate of the hit position. The hits on the sensors (other than
those in the module under study) are fitted to produce tracks. These
tracks are extrapolated to obtain the track intercept point in the
sensor under study.  The unbiased residual is then defined as the
perpendicular distance between the track intercept point and the line
parallel to the strip at the observed hit position. Consequently,
these residuals are sensitive to sensor misalignments perpendicular
to the strip direction.
Since the direction of the strips changes as a function of the
azimuthal angle the residuals thus have a well defined sensitivity to
translational misalignments in $\x$ and $\y$ (see Fig.~\ref
{fig:STEP0_sketch}).

It can easily be shown that the relation between residuals ($\epsilon_
{\R/\Phi}$) and misalignments in the sensor plane ($\Delta_{\x},
\Delta_{\y},\Delta_{\gamma}$) is given by

\begin{equation}
\left\{
\begin{array}{lcllll}
\epsilon_{\R}    & = & - \Delta_{\x}\cos\phi_{track} & + \Delta_{\y}
\sin\phi_{track} & & (\R\mathrm{~sensor})\\
\epsilon_{\Phi} & = & + \Delta_{\x}\sin\phi_{track} & + \Delta_{\y}
\cos\phi_{track} & + \Delta_{\gamma} r_{track} & (\Phi\mathrm{~sensor})
\end{array}
\right. ,
\label{eqn:step0}
\end{equation}
where $\phi_{track}$ is the azimuthal angle of the extrapolated track
position.
$\Delta_{\gamma}$ describes a misalignment in the form of a rotation
around
the $z$ axis, which translates into a shift in $\phi$ by
multiplication with the radial co-ordinate of the extrapolated track
in the sensor plane. As this does not contain any $\phi$ dependence
it is sufficient to leave it as a free parameter in the fit.

In practice the alignment constants are determined by an iterative
fitting algorithm.
For each iteration, the unbiased residuals for the $\R$ and $\Phi$
sensor are plotted against the azimuthal angle. The misalignments are
determined through a fit to a binned distribution of the mean
residual values as function of $\phi$.

An additional complication arises from the fact that the strips on
the VELO $\Phi$ sensors are not exact radial lines. The $\Phi$
sensors are divided into an inner and an outer region within which
the strips are tilted through a stereo angle of $20^\circ$ (inner)
and $-10.35^\circ$ (outer).
This can be compensated for by making the replacement in Eq.~\ref
{eqn:step0}
$$\phi\rightarrow\phi'=\phi_{min}+\beta, $$
where $\beta$ is the stereo angle and $\phi_{min}$ is the $\phi$
coordinate at the minimal radius of the strip. In addition to
preserving the simple relation between the residuals and the
alignment constants, this transformation also allows the fit to be
made to data from the inner and outer region of the $\Phi$ sensor
simultaneously.

This method is primarily used to determine the relative \x\ and \y\ translations of the sensors in a module.
However, to improve the fit convergence, the value for $\Delta_{\gamma}$, the rotation around the $z$-axis of the $\Phi$ sensor, is also determined.
This is done using an equivalent approach as above, however now exploiting the distribution of residuals as a function of the track radial co-ordinate.
The value for $\Delta_{\gamma}$ is then determined as the slope of fit with a linear function.
However, this  relative \z\ rotation of the modules will be extracted to a higher precision using the technique described below in step two of the
alignment procedure.

The fitted tracks used in the procedure rely upon the current
estimate of the alignment constants of the sensors. Hence it is
necessary to iterate this procedure, updating the fitted tracks as
the alignment constants are improved.

As the unbiased residuals are determined from a track fitted while excluding hits on both sensors under study, the influence of module to module misalignments vanishes to first order when determining the relative translational misalignment of the sensors.
The only misalignments that cannot be determined with this method are common sensor to sensor misalignments of all modules, and sensor to sensor misalignments that have a linear dependency on $z$.

The results of applying this first step of the alignment procedure to
simulation events are given in section \ref{sec:res1}.

\subsection{Relative Module Alignment}

The second step of the alignment procedure is to perform the relative
alignment of the 21 VELO modules within one detector-half. This is
based on a non-iterative method using a matrix inversion technique to
minimize a $\chi^{2}$ function.

The $\chi^{2}$ is produced from the residuals between the tracks and
the measured clusters. The measured residuals are expressed as a
linear combination of the track parameters and the alignment
constants. Each straight line track has four parameters (two slopes
and two intercepts) denoted $n_{local}$. The total number of
translational and rotational degrees of freedom of all of the modules
is $n_{global}$ alignment constants. The track fitting and alignment
problem can then be expressed as a system of $n_{total}$ equations,
where $n_{total}$ is given by $$n_{total}~=~n_{local}\cdot n_{tracks}~
+~n_{global}.$$

Clearly, the size of the system scales with the number of tracks used
in the alignment. As discussed later, about 20,000 tracks are
necessary to obtain  a detector-half alignment of the required
accuracy. Hence, this implies solving a system of over 80,000
equations, which is a computationally challenging task. However, this
problem can be reduced to the size of $n_{global}$ (around 100 for
the VELO) using the technique of matrix inversion by partition,
performed by the program Millepede{\footnote{a {\tt C++} translation
of the {\tt FORTRAN} Millepede program has been implemented by one of
the authors of this paper~\cite{bib:Vir-05}.}}~\cite{bib:Blo-07}. This algorithm has already been successfully used in many particle physics experiments (see for example~\cite{bib:Her-07,bib:Zeu-07})

As stated previously, VELO $\R$-sensor strips are semi-circular, thus making $\R$-sensor hit information non-linear. However, in order to use the Millepede approach it is mandatory to establish a linear relation between the hit residuals and the misalignments. Here, this is obtained by producing space-points at the VELO module level as described below.  This is the natural choice for the VELO as a module containing an \R\ and a $\Phi$ sensor is a independent mechanical object whose internal alignment is expected to be relatively stable.

A particle
passing through a VELO module (which is assumed to have its \R\ and $
\Phi$ sensors already internally aligned) gives a non-linear set of
coordinates: $(\rvar,\z(\R_{sensor}))$ and $(\phi,\z(\Phi_{sensor}))
$. This system is transformed into an $(\x,\y,\z)$ space-point by
projecting the $\phi$ information onto the $\R$ sensor. As the $\phi$ co-
ordinate of the particle can change slightly in the gap between the \R\
sensor and the $\Phi$ sensor, a correction based on the currently
estimated track slopes is applied. Since the correction is small, and
the track slope estimates are not significantly biased by the
expected module misalignments, the non-linear effects introduced by
this procedure are negligible.

An $(r,\phi)$ cluster pair is thus transformed into an $(\x,\y,\z)$
space-point as follows:
$$
\left\{
\begin{array}{lcl}
\x & = & \rvar\cdot \cos(\phi_{corr}) \\
\y & = & \rvar\cdot \sin(\phi_{corr}) \\
\z & = & \z(\R_{sensor})
\end{array}
\right.
$$
with $$\phi_{corr} = \phi + \phi_{track}(\R_{sensor}) - \phi_{track}
(\Phi_{sensor}).$$

The linear relation between the residuals and the alignment
parameters for the VELO modules, which is derived in Ref.~\cite
{bib:Vir-05}, is given by:
\begin{equation}
\left\{
\begin{array}{lclll}
\epsilon_{\x} & = & - \Delta_{\x} & + \y\cdot\Delta_{\gamma} & + a
\cdot(\Delta_{\z} + \x\cdot\Delta_{\beta} + \y\cdot\Delta_{\alpha})\\
\epsilon_{\y} & = & - \Delta_{\y} & - \x\cdot\Delta_{\gamma} & + c
\cdot(\Delta_{\z} + \x\cdot\Delta_{\beta} + \y\cdot\Delta_{\alpha})
\end{array}
\right.
\label{eqn:step1}
\end{equation}
where $\x$, $\y$, $\z$ represents the measurement values, and $a$, $c
$ are the slopes in the $\X\Z$, $\Y\Z$ planes respectively of the
track considered. The Millepede technique is then used to
simultaneously extract the alignment constants ($\Delta_{\x},~\Delta_
{\y},~\Delta_{\z},~\Delta_{\alpha},~\Delta_{\beta}$ and $\Delta_
{\gamma}$) of each of the 21 VELO modules.

An important component of the alignment procedure is the selection of
the track sample used. In order to ensure an optimal population of
the final matrix, a mixture of tracks coming from the primary
interaction point and a complementary set of tracks from the beam
halo or beam-gas interactions will be used. A specific pattern
recognition algorithm has been developed in order to select these
events~\cite{bib:Las-07}.

The `weak-modes', \ie\ deformations which are difficult --- if not
impossible --- to unfold with tracks, are extensively constrained.
Two categories of `weak-modes' have been considered.

\begin{figure}[h!]
\centering
\includegraphics[width=0.65\textwidth]{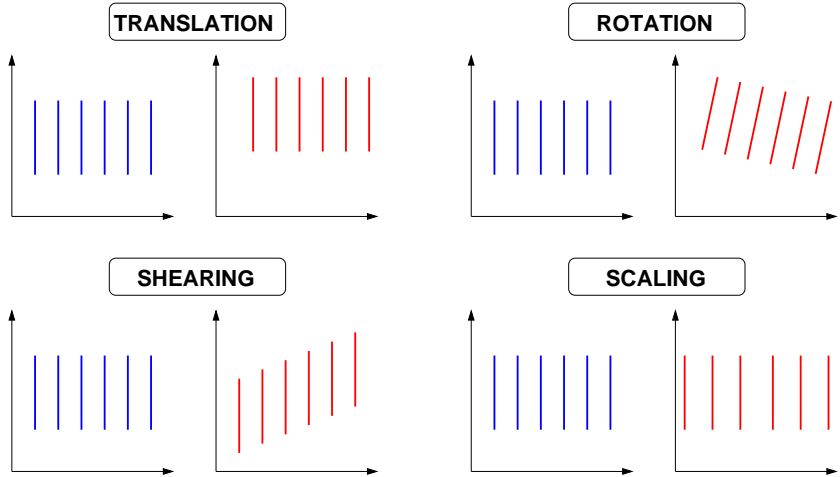}
\caption{Linear deformations impossible to constrain via VELO module
alignment.}
\label{fig:Deformations}
\end{figure}

Firstly, track residuals are insensitive to linear transformations of
the whole detector-half (\eg\ translation along an axis, or rotation
around an axis). The different possible effects are illustrated
pictorially in Fig.~\ref{fig:Deformations}. In three dimensions, this
leads to 12 possible global deformations. However, due to the
structure of the VELO, we can neglect three of them at first order:
shearing in the XY plane, and scaling of the X and Y axes. In
addition, due to the VELO geometry, it will be difficult to
distinguish XZ and YZ shearings from Y and X rotations. We choose to
constrain only the XZ and YZ shearings, as they are easier to add
into our linear system.

Hence we constrain 7 possible deformations: Z axis rotation, X, Y, Z
translations, XZ and YZ shearings, and Z axis scaling. These degrees
of freedom are fixed in order to prevent these movements occurring
during the module alignment. Within the Millepede framework, this is
achieved through Lagrange multipliers, \ie\ through the addition of
constraint equations to the global system.

The second class of `weak modes' are those that do affect the track residuals but to which there is greatly reduced sensitivity. This is, for example, the case for module rotations around the \X-axis, or around the \Z-axis with an angle proportional to \Z. Even with the best possible track sample the effect on the $\chi^{2}$ of these modes is comparatively weak, hence they are also weakly constrained by any alignment procedure. Using the survey information as a starting point for the alignment will provide a strong constraint against those special modes. In addition to that, and in order to avoid non-physical variations during the alignment procedure, each alignment constant is controlled: `penalty-terms' are added to the $\chi^{2}$ to minimize. These extra-terms prevent the variation of the alignment parameters (w.r.t. the initial survey value) to be significantly larger than the metrology precision.

The results of applying this second step of the alignment procedure
to simulation events are given in section \ref{sec:res2}.

\subsection{Detector-halves Alignment}

The final step of the alignment procedure is to perform the relative
alignment of the two detector-halves. Clearly, the module space-point
residuals within a detector-half are not sensitive to the
misalignment of the whole half. Thus, observables connecting the two
detector-halves have to be used.

Tracks that pass through both VELO detector-halves, referred to here
as overlap tracks, provide a powerful constraint for the relative
positioning of the two detector-halves. These tracks are required to
have at least one space-point (\ie\ an $(\rvar,\phi)$ cluster pair)
in each VELO detector-half. The VELO was specifically designed to
obtain such a class of tracks: the modules from the VELO left and
right-halves are offset in $\z$ and interlace by a small distance
when closed. The overlapping area between the two VELO detector-
halves, when fully inserted, corresponds to 2\% of the active surface
area of the sensors.

Having selected these overlap tracks, the alignment proceeds by
residual minimization using the same technique as the module
alignment. An equivalent of Eqn.~\ref{eqn:step1} is thus required in
terms of the relative alignment constants of the two detector-halves.
Assuming that the alignment of the modules in each detector-half has
been already corrected, then the equation for the residuals becomes~
(see Ref.~\cite{bib:Vir-07}):
$$
\left\{
\begin{array}{ccl}
\epsilon_{\x} & = & - \Delta^B_\x + \z_0^B\cdot \Delta^B_{\beta} + \y
\cdot\Delta^B_{\gamma} + a\cdot (\Delta^B_{\z}+\x\cdot\Delta^B_{\beta}
+\y\cdot\Delta^B_{\alpha})\\
\epsilon_{\y} & = & - \Delta^B_\y + \z_0^B\cdot \Delta^B_{\alpha} - \x
\cdot\Delta^B_{\gamma} + c\cdot (\Delta^B_{\z}+\x\cdot\Delta^B_{\beta}
+\y\cdot\Delta^B_{\alpha})\\
\end{array}
\right. ,
$$
where $\Delta^B_\x,~\Delta^B_\y,~\Delta^B_\z,~\Delta^B_{\alpha},~
\Delta^B_{\beta},~\Delta^B_{\gamma}$ are the detector-half degrees of
freedom (position of one half with respect to the other), and $\z_0^B
$ the $\z$ position in the detector-half frame of the module in which
the space-point hit was recorded.

The matrix inversion technique then allows these six relative
alignment constants of the detector-halves to be determined, assuming
the VELO is fully inserted for physics data taking.

However, during insertion and during the commissioning phase an
alignment may be required with the VELO in the retracted position. In
this case the rate of overlap tracks becomes very small and an
alternative technique is required. The detector-half alignment can
then be performed using vertices as constraints. Tracks from the
primary interaction point can be used if available or beam-gas
interactions occurring in the VELO vacuum tank. By fitting for the
vertex separately inside each of the two detector-halves, one can
obtain the misalignment between the two halves.

Assuming that one box is fixed (\ie\ using it to define the co-
ordinate system) the reference vertex position can be determined from
this half. The vertex position found using the tracks passing through
the other detector-half is related to the misalignment between the
two halves via the following relations:
$$
\left\{
\begin{array}{lcl}
v_\x^B & = & b_i + a_i\cdot v_\z^B + \Delta^B_\x - a_i\cdot \Delta^B_
\z - v_\z^B\cdot\Delta^B_{\beta} \\
v_\y^B & = & d_i + c_i\cdot v_\z^B + \Delta^B_\y - c_i\cdot \Delta^B_
\z - v_\z^B\cdot\Delta^B_{\alpha}
\end{array}
\right. ,
$$
where $a_i, b_i, c_i, d_i$ are the parameters of the $i^{th}$ track
used to fit the vertex, $(v_\x^B,v_\y^B,v_\z^B)$ the vertex position
in the `moving' half, and $\Delta^B_i$ denote the detector-half
degrees of freedom (position of one half with respect to the other).
It is easy to relate this formulation to the residual formalism used
previously. Indeed, the residual in this case will be the differences
between the vertices positions fitted within the two halves:
$$
\left\{
\begin{array}{ccccl}
v^{B}_\x & = & v^{ref}_\x + \epsilon_{v_\x} \\
v^{B}_\y & = & v^{ref}_\y + \epsilon_{v_\y} \\
v^{B}_\z & = & v^{ref}_\z + \epsilon_{v_\z} \\
\end{array}
\right. .
$$
where $(v_\x^{ref},v_\y^{ref},v_\z^{ref})$ defines the vertex
position in the fixed half, \ie\ the reference value.

Having determined these relations, the same technique can again be
used to determine the alignment constants: even though we are now
using vertices rather than track residuals the Millepede framework is
still applicable as it does not depend on the object that is being
fitted.

More details on this novel technique of alignment with vertices can
be found in Ref.~\cite{bib:Vir-07}. The vertex fitting method can
also be used to determine the VELO detector-halves' positions with
respect to the beam, when using vertices from the primary interaction
point.

The results of applying this final step of the alignment procedure to
simulation events are given in section \ref{sec:res3}.

\section{Simulation Studies}
\label{sec:MC_results}

A simulation of 200 samples of 25,000 events each has been produced
and propagated through the LHCb software. Each sample, which
comprises a mixture of 5,000 minimum bias events ($\approx$~100,000
tracks from primary vertex interactions) plus 20,000 beam-halo like
events, was produced with a different set of alignment constants. The
misalignment values have been randomly chosen within a Gaussian
distribution centered on zero and with resolutions based on
construction and survey accuracies (defined in Ref.~\cite
{bib:Vir-07}). All the module and detector-half degrees of freedom
have been misaligned. The initial step, the relative sensor
alignment, was tested using different event samples which are
described in the first part of this section.

\subsection{Relative Sensor Alignment Results}

\label{sec:res1}

The sensor alignment method has been tested with 10 samples of
randomly generated misalignments. All sensors have been misaligned individually, thus generating a scenario equivalent to simultaneous module to module and sensor to sensor misalignments. Each of the 10 samples consists of 20,000
tracks with small slopes, thus passing through all sensors of one
VELO-half and evenly distributed across the sensor surface. Typically
three iterations of the alignment procedure are required to obtain
the best resolution.

\begin{figure}[h!]
\begin{center}
\begin{minipage}[t]{6.5cm}
     \resizebox{6.5cm}{!}{\includegraphics{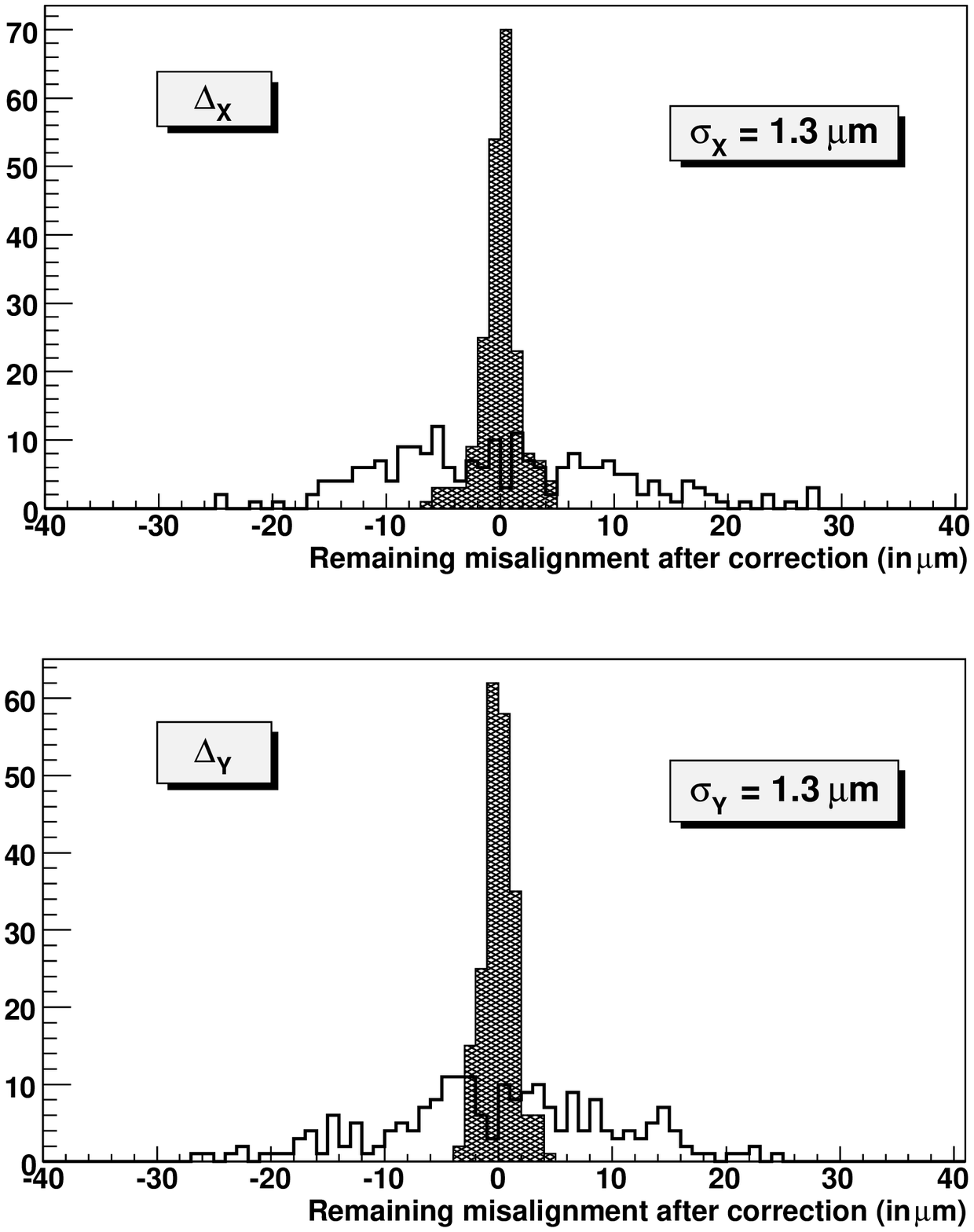}}
    \caption{Misalignment values before ($\square$), and after ($
\blacksquare$) sensor relative alignment.}
    \label{fig:STEP0_result}
\end{minipage}
\hfill
\begin{minipage}[t]{6.5cm}
     \resizebox{6.5cm}{!}{\includegraphics{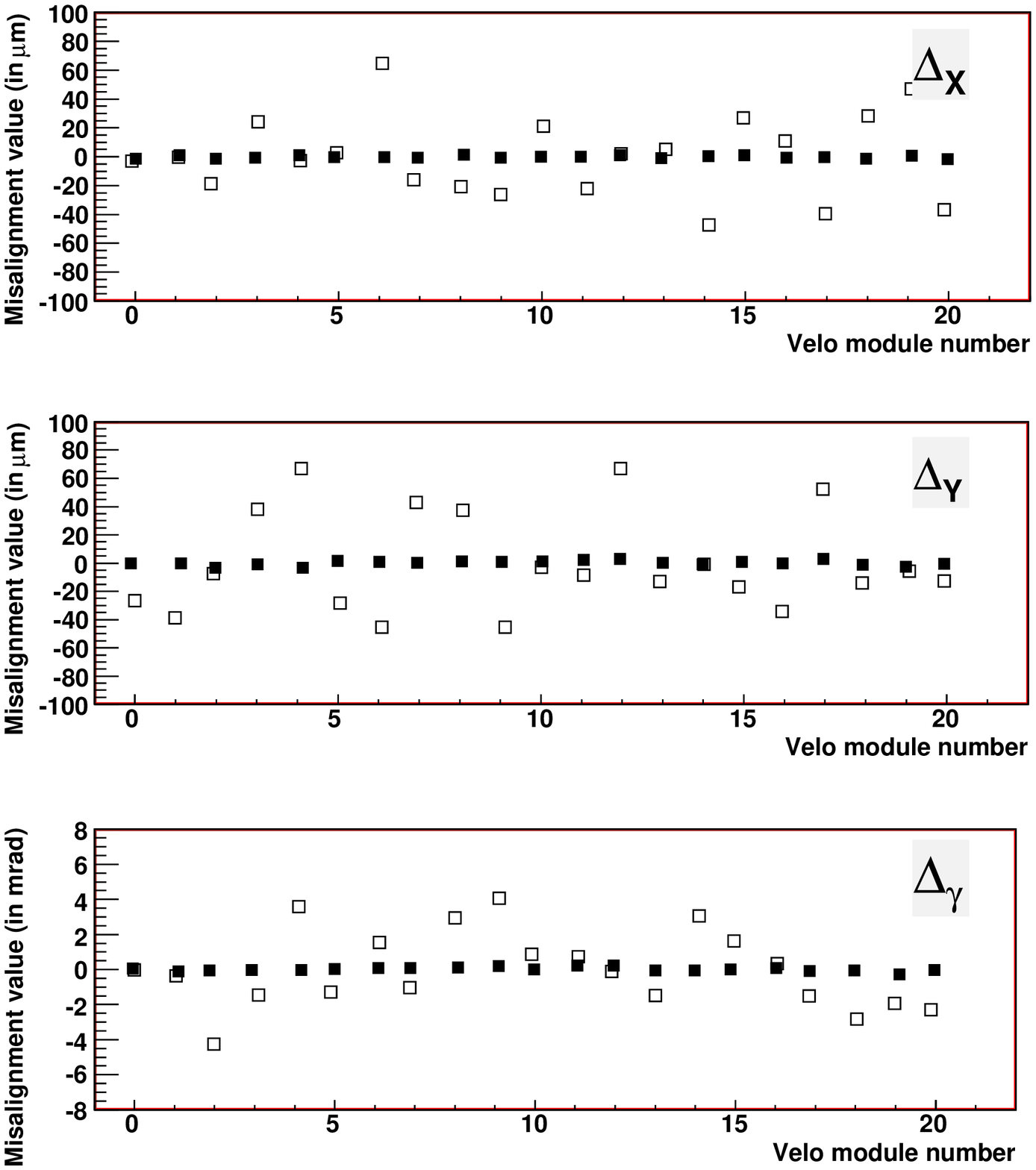}}
     \caption{Misalignment values in one detector-half before ($
\square$), and after ($\blacksquare$) internal alignment.}
     \label{fig:STEP1_stations}
\end{minipage}
\end{center}
\end{figure}

Fig.~\ref{fig:STEP0_result} shows the generated and the remaining
misalignments after all iterations. The resolution on the relative $\x
$ and $\y$ translation of the sensors of one module is $1.3~\mathrm
{\mum}$, \ie\ a significant improvement over the survey precision.
The performance of this algorithm has also been demonstrated with
beam test data and is reported in Ref.~\cite{bib:NIM-07}.

\subsection{Relative Module Alignment Results}

\label{sec:res2}

The internal alignment of the modules in each detector-half is
primarily sensitive to translations of the modules in the $\X$ and $\Y
$ directions and rotations around the $\Z$-axis. Results on the
alignment of all modules in a detector-half, for one particular
sample, are presented in Fig.~\ref{fig:STEP1_stations}. In Fig.~\ref
{fig:STEP1_fit} the alignment constants for 200 event samples, before
and after correction, are shown. Resolutions on the \X\ and \Y\
translation alignment parameters of $1.1~\mum$ and on rotations
around the $z$-axis of $0.12~\mrad$ are obtained. About 20,000 tracks
per detector-half were needed to obtain this accuracy. The alignment
resolutions are well below the intrinsic detector hit resolution. The
performance of this algorithm has also been demonstrated with beam
test data and is reported in Ref.~\cite{bib:NIM-07}.

Concerning the `non-linear' degrees of freedom, the observed
sensitivity is as expected worse than for the other parameters.
However some results were obtained for the modules which are close to
the interaction region, \ie\ where track slopes are larger.
Restricting the study to these stations (1 to 14), one obtains a
reasonable sensitivity to $\Delta_\z$ ($28~\mum$) and a fair
sensitivity to $\Delta_{\alpha}$ and $\Delta_{\beta}$ (0.8~$\mrad$
and 1.1~$\mrad$ respectively). This sensitivity is worse than
the survey precision, but will provide a cross-check of this survey information.

As this algorithm is in general run independently of the relative sensor alignment algorithms its performance has been evaluated separately. In the presence of relative misalignments of the sensors on a given module the module's position will be aligned to the average position of the two sensors\footnote{This requires a track sample with a sufficiently flat distribution in $\phi$ which is given for the samples used for VELO alignment.}.

\subsection{Detector-halves Alignment Results}

\label{sec:res3}

Although the three alignment steps can be performed independently, in
practice it is expected that steps two and three will be run
consecutively. Hence, the results presented in this section are for
the realistic case of performing both of these alignment steps on
misaligned samples. The tracks are refitted after the module
alignment procedure in order to update the track parameters. The
results presented here have been obtained with about 300 overlap tracks.

\begin{figure}[h!]
\begin{center}
\begin{minipage}[t]{6.5cm}
     \resizebox{6.3cm}{!}{\includegraphics{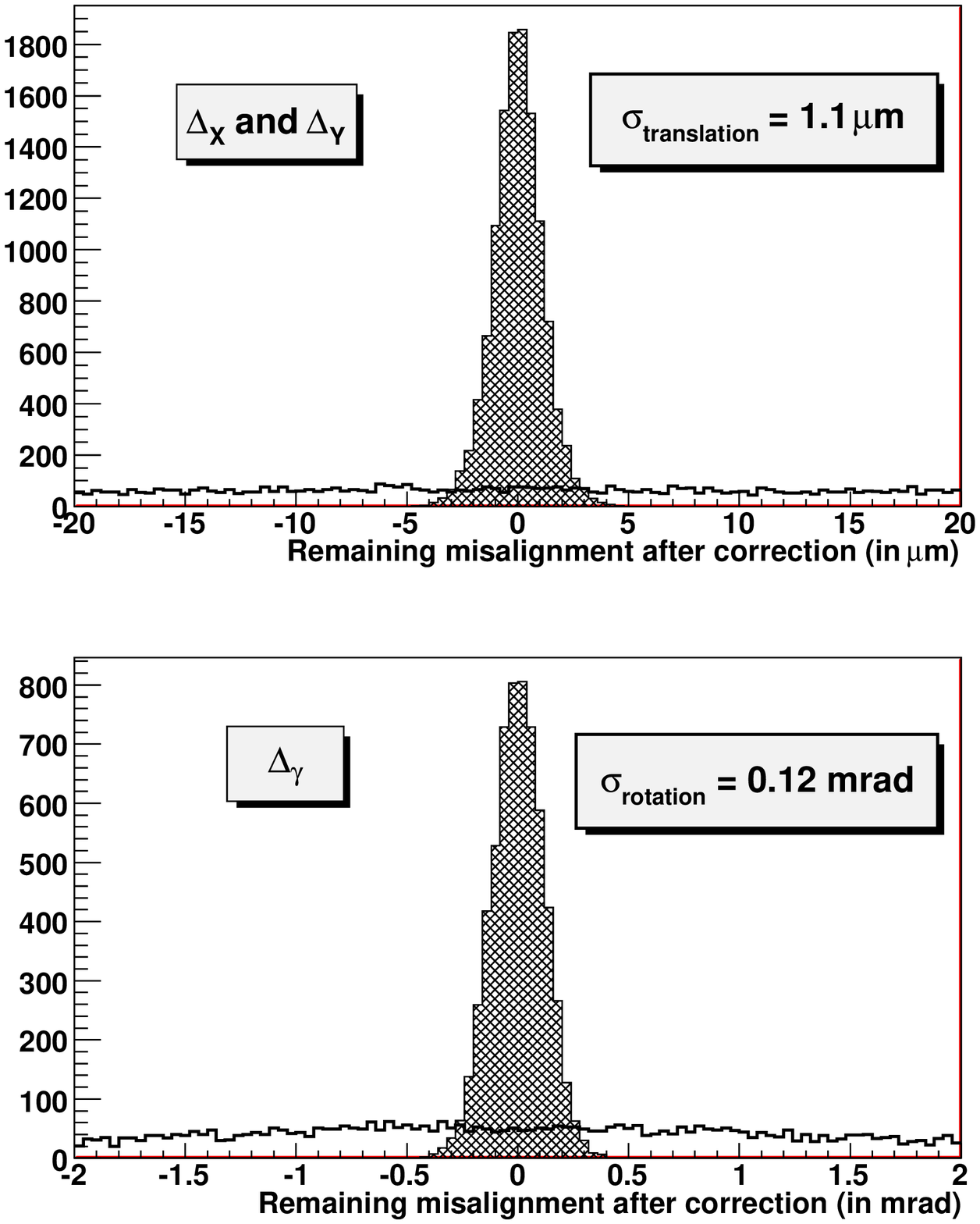}}
     \caption{Misalignment values before ($\square$), and after ($
\blacksquare$) internal alignment, for all configurations and all
stations.}
     \label{fig:STEP1_fit}
\end{minipage}
\hfill
\begin{minipage}[t]{6.5cm}
     \resizebox{6.5cm}{!}{\includegraphics{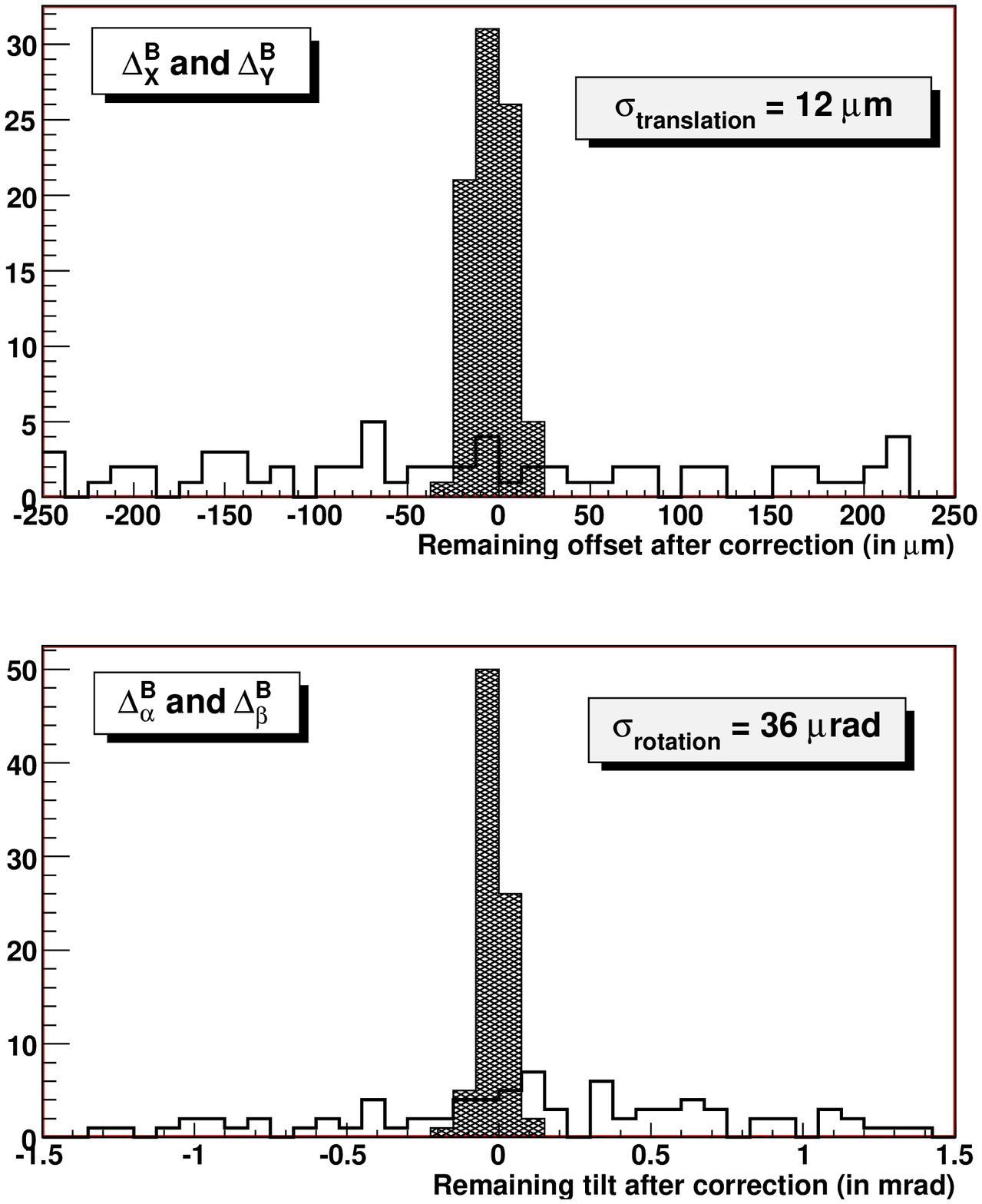}}
     \caption{Misalignment values before ($\square$), and after ($
\blacksquare$) detector halves alignment, for all configurations.}
     \label{fig:STEP2_fit}
\end{minipage}
\end{center}
\end{figure}

The results of the study are presented on Fig.~\ref{fig:STEP2_fit}.
The resolution on the X and Y translation alignment parameters is $12~
\mum$ for x and y translations, and the resolution on the x and y
tilts is $36~\murad$.

As in the case of the module alignment, some of the degrees of
freedom are more difficult to constrain. In the detector-half
alignment these weakly constrained motions are the ones related to
the Z-axis: rotation around and translation along. The relative
rotation around Z between the two detector-halves is constrained
using the overlap tracks. Translations along Z are estimated through
the vertex fitting technique. The vertices are fitted separately from
tracks in each detector-half and the misalignment determined by
comparing the Z positions, from this a 40~$\mum$ resolution was
obtained.

\section{Conclusion}
\label{sec:conclusion}

A software alignment method for the LHCb vertex locator has been
developed. This procedure performs the alignment in three steps, all
using track residuals. First, a relative alignment of the $\R$ and $
\Phi$ sensors within each VELO module is performed. This is followed
by an internal alignment of the VELO modules within each detector-
half. Thirdly, since the detector-halves are moved between each LHC
fill, a final step is required in order to align the detector-halves
with respect to each other, thus providing a fully internally aligned
VELO.

Due to the VELO module design ($\R$ and $\Phi$ sensors are glued onto
the same hybrid and precisely surveyed), it is foreseen to perform
the first stage with a much lower frequency than the two other ones
and to use it for the annual data reprocessing. The alignment of the
relative sensor position achieves a precision of $1.3~\mum$ for $\x$
and $\y$ translations.

The final two stages of VELO alignment strategy use the Millepede
program, which enables the alignment to be performed in only one
pass; this is to be compared with classic minimization methods which
require many iterations to provide their result. This technique
allows the processing of module and VELO-half alignment within a few
minutes on a single CPU\footnote{1 CPU = 1000 SpecInt2000 units},
assuming that an appropriate data sample is available.

For the internal alignment of the modules in a detector-half a $1.1~
\mum$ precision has been achieved on the relevant translational
degrees of freedom (\ie\ along $\X$ and $\Y$ axes), and a $0.1~\mrad$
accuracy on the rotation around the $\Z$-axis.  These values are
well below the expected detector intrinsic resolution. Future work
will be undertaken to enable an efficient selection of the required
tracks in the LHCb trigger.

The results of the detector-half alignment procedure show that the
tracks that pass through the overlap region between the two VELO
detector-halves provide a very strong constraint. With only a few
hundred tracks accuracies of $12~\mum$ for x and y translations, and
$36~\murad$ for $\X$ and $\Y$ tilts, are obtained. These results are
well within the system requirements.

This paper has described the algorithms developed to perform the
alignment of the LHCb VELO and demonstrated with simulation samples
that this approach provides the performance required so that
misalignment effects will not adversely affect the LHCb physics
programme.

\end{document}